\documentclass[12pt,a4paper,english,superscriptaddress,aps,nofootinbib]{revtex4}
\usepackage[utf8]{inputenc}
\usepackage[T1]{fontenc}
\usepackage{amsmath,amssymb,graphicx}
\makeatletter
\usepackage{babel}
\usepackage[active]{srcltx}
\usepackage{graphicx,color}
\usepackage{changebar}
\usepackage{hyperref}
\usepackage[T1]{fontenc}
\usepackage{esint}
\usepackage{multirow}
\usepackage{dsfont}
\usepackage{ae}
\usepackage{amsmath}
\usepackage{multirow}
\usepackage{braket}
\usepackage{mathtools}
\usepackage{slashed}
\usepackage{empheq}
\usepackage{multirow}
\usepackage{graphicx}
\usepackage{amsfonts}
\usepackage{amsmath}

\begin{document}
    \title{On the singular position-dependent mass}
    
\author{F. C. E. Lima}
\email{cleiton.estevao@fisica.ufc.br}
\affiliation{Departamento de F\'{i}sica, Universidade Federal do Maranh\~{a}o, S\~{a}o Lu\'{i}s, MA,  Brazil.}
\affiliation{Departamento de F\'{\i}sica, Universidade Federal do Cear\'{a}, Fortaleza, CE, Brazil.}

\author{F. M. Belchior}
\email{belchior@fisica.ufc.br}
\affiliation{Departamento de F\'{\i}sica, Universidade Federal do Cear\'{a}, Fortaleza, CE, Brazil.}

\author{C. A. S. Almeida}
\email{carlos@fisica.ufc.br}
\affiliation{Departamento de F\'{\i}sica, Universidade Federal do Cear\'{a}, Fortaleza, CE, Brazil.}
\affiliation{Departamento de F\'{i}sica Te\'{o}rica, Universidad de Valencia, Burjassot, Valencia, Spain.}

\begin{abstract} 
\vspace{0.5cm}
\noindent \textbf{Abstract:} 
Revisiting the issue associated with Position-Dependent Mass (PDM), we reaffirm that the appropriate framework for addressing a generic PDM is the symmetrization proposed by BenDaniel and Duke. To accomplish this result adopts the effective mass Hamiltonian proposed by von Roos, corrected by a symmetrized kinematic term. After verifying the appropriate ordering to approach the PDM issue, one investigates a crystalline lattice with a defect described by a singular PDM. The singular mass profile proves intriguing as it yields an atom's cluster in the neighborhood of the singularity. Considering that a restoring force acts on the atoms, one notes that the confluent Heun function describes the quantum states.  Furthermore, one highlights that when the effective mass distribution tends to a constant profile, we recover a system similar to the harmonic oscillator.

\vspace{0.35cm}
\end{abstract}

\maketitle

\thispagestyle{empty}
\newpage


\section{Introduction}

In 1926, Erwin Schr\"{o}dinger proposed the equation that describes quantum mechanical systems \cite{Schrodinger}. Since then, studies on the exact solutions of quantum systems have attracted significant attention \cite{Yu,Schiff,Dirac}. An essential method for studying these systems is the series expansion method \cite{Ince}, which has shown to be appropriate in studies on quantum system solutions. For instance, one employs this approach in the quantum description of the harmonic oscillator \cite{Griffiths}, the hydrogen atom \cite{Griffiths}, and other relevant physical systems described by several types of interactions \cite{Nunez,Wang,Almeida0,Sun,Chen1,Chen2,IWang}.

Essentially, one formulates Schr\"{o}dinger's theory considering particles with constant mass \cite{Schrodinger}. However, in addressing solid-state physics systems, it becomes evident that for a theoretical understanding of transport phenomena in semiconductors, the chemical composition of the material is position-dependent \cite{Slater,Luttinger}. Thus, the PDM concept emerges when a periodic field acts on a particle, i.e., an electron. In this scenario, one promotes the particle's mass to a tensorial entity concerning unperturbed band structure \cite{Luttinger}. The electronic wave packet that emerges in the energy band constitutes the entity correlated with the PDM. This wave packet follows the Wannier-Slater theorem \cite{Slater,Wannier}, whose expression resembles Schr\"{o}dinger's equation, i.e.,
\begin{align}
    [E(k)(-i\nabla+V(r))]F(r,t)=i\hbar \dot{F}(r,t),
\end{align}
where $F(r,t)$ is the envelope function, $E(k)$ the dispersion energy, $k$ the crystal momentum, and $V(r)$ the potential due to external sources (impurities). Nonetheless, one highlights that considering a many-body Hamiltonian approximation to an electron, it is possible to derive a Schr\"{o}dinger-like equation with PDM. In this context, the PDM concept has been the subject of intensive study in several systems with effective mass \cite{Mustafa,Costa,Zare,Ghafourian,Pourali}. Furthermore, it is noteworthy that significant recent technological and theoretical advances have motivated considerable investigation into systems with PDM. Such interest arises from the crucial role played by this property, which enables the analysis of semiconductor impurities \cite{Slater,Luttinger,Wannier}, nonlinear optical phenomena \cite{Guo,Zhang}, and quantum-relativistic problems \cite{ILima0}. In addition to these applications, PDM problems are extensively used in studies concerning the influence of effective mass on polarons confined in parabolic wells \cite{RR1}, systems with complex optical potentials \cite{RR2}, as well as the quantum-mechanical description of systems subject to hyperbolic interactions \cite{RR3,RR4}, among other \cite{RR5,RR6,RR8,RR9,RR10,RR11,RR12}. Although the study of problems involving PDM is a well-established field, we propose, for the first time, that a crystalline lattice, whose atoms inherently vibrate under the influence of a harmonic potential (see Ref. \cite{Kittel}), can exhibit singular characteristics when subjected to parallel stress induced by impurities, such as electromagnetic fields. This situation leads to the emergence of the singular mass profile, derived from multipolar contributions, as will be detailed in Eq. \eqref{Eq6}. In this framework, this work aims to investigate the quantum-mechanical description of the system resulting from the clustering of particles caused by such impurities. Although several studies have addressed problems related to singular PDM, we believe this is the first to explore the configuration of lattice defects using the PDM formalism.

In the relativistic regime, the PDM problems do not exhibit ambiguities \cite{ILima0}. Nevertheless, the symmetrization ambiguity of the kinematic operator arises in the non-relativistic regime. Until now, there has been no consensus on the form of the symmetrization without ambiguity of the kinetic energy operator \cite{Cavalcante}. In Ref. \cite{Cavalcante}, one displays an attempt to resolve the problem of the symmetrization of the kinematic operator, i.e., the authors adopt a relativistic theory taking a non-relativistic bound using the Foldy-Wouthuysen approach \cite{FW}. Considering this approach, the authors showed that the ordering proposed by Li and Kuhn \cite{LiKuhn} is consistent with the relativistic PDM theory. However, this remains an open problem in the literature. Thus, other orderings of the kinematic operator have emerged and describe abrupt semiconductor heterojunctions and other materials. In the literature, one finds four widely used orderings to symmetrize the kinematic operator and bypass the ambiguity problem, viz., Li-Kuhn \cite{LiKuhn}, BenDaniel-Duke \cite{BenDaniel}, Zhu-Kroemer \cite{ZhuKroemer}, and Gora-Williams \cite{Gora}.

We outline our research's purpose into two distinct parts. The first purpose is to propose a suitable non-relativistic operator for a generic PDM. Through this analysis, one demonstrates that an unambiguous kinematic ordering, which respects Heisenberg's Uncertainty Principle and is appropriate for addressing PDM problems while preserving the von Roos ordering \cite{Roos}, i.e.,
\begin{align}
    \hat{K}=\frac{1}{4}\{m^\alpha(\text{r})\,\hat{p}\,m^\beta(\textbf{r})\hat{p}\,m^{\gamma}(\textbf{r})+m^\gamma(\text{r})\,\hat{p}\,m^\beta(\textbf{r})\hat{p}\,m^{\alpha}(\textbf{r})\},
\end{align}
is described by the BenDaniel and Duke ordering \cite{BenDaniel}. This ordering is responsible for describing semiconductor heterostructures. The second purpose of this article is the quantum-mechanical description of a crystalline lattice with a defect outlined by a singular PDM. The singularity of the PDM is of interest due to its ability to modify the equilibrium distances of the lattice sites, generating a particle's cluster near the singularity. Thus, our study may establish the first step towards understanding the quantum behavior of crystalline lattices with defects in mass (or particle) distribution.

We structure the manuscript into four distinct sections. Firstly, we discuss and propose the most appropriate way to describe the kinetic energy operator in Sec. \ref{secII}. In Sec. \ref{secIII}, one adopts a singular mass profile and the assumption that the lattice sites are subject to a restoring force. Thus, one performs the quantum description of the system. Finally, we present our remarks and conclusions in Sec. \ref{secIV}.


\section{The Generalized Schr\"{o}dinger theory}\label{secII}

Let us start our analysis by considering the non-relativistic theory in the most general context possible, i.e., adopting the position-dependent mass problem.  In this scenario, one assumes a Hamiltonian similar to the proposed by von Roos \cite{Roos}, namely,
\begin{align}\label{Eq1}
    \hat{K}=\frac{1}{4}\{[m^{-1}(\textbf{r})\,\hat{p}^2+\hat{p}^2\,m^{-1}(\textbf{r})]+m^\alpha(\text{r})\,\hat{p}\,m^\beta(\textbf{r})\hat{p}\,m^{\gamma}(\textbf{r})+m^\gamma(\text{r})\,\hat{p}\,m^\beta(\textbf{r})\hat{p}\,m^{\alpha}(\textbf{r})\}, 
\end{align}
The parameters $\alpha$, $\beta$, and $\gamma$ obey to the constraint $\alpha+\beta+\gamma=-1$. Furthermore, $m(\textbf{r})$ denotes the position-dependent mass, $\hat{p}=-i\hbar\nabla$ the momentum operator, and $\hbar$ the reduced Planck constant. The first term of Eq. (\ref{Eq1}) is implemented to account for the usual symmetrized or Weyl-ordered operator (i.e., $\alpha=\beta=0$) \cite{Almeida,Lee}.

In one-dimensional space, properly commuting the momentum operator $\hat{p}=-i\hbar (d/dx)$ to the right, one has the effective operator
\begin{align}\label{Eq2}
    \hat{K}=\frac{1}{2m(x)}\hat{p}^2+\frac{i\hbar}{2}\frac{m'(x)}{m^2(x)}\hat{p}+U_k (x), 
\end{align}
where
\begin{align}\label{Eq3}
    U_k(x)=-\frac{\hbar^2}{4m^3(x)}\bigg[(\alpha+\gamma-1)\frac{m(x)}{2}\frac{d^2m(x)}{dx^2}+(1-\alpha \gamma-\alpha-\gamma)\bigg(\frac{dm(x)}{dx}\bigg)^2 \bigg].
\end{align}
The potential $U_k$ originates from the kinematic term, in which different combinations of the von Roos parameters lead us to several interactions. One can bypass this problem by imposing the condition
\begin{align}\label{Eq4}
    \alpha+\gamma=\alpha\gamma+\alpha+\gamma=1.
\end{align}
The equation (\ref{Eq4}) leads us to $\alpha=0$ and $\gamma=1$, or $\alpha=1$ and $\gamma=0$. These conditions result in a kinetic operator free from ambiguities related to Heisenberg's uncertainty principle. Note that condition (\ref{Eq4}) excludes the possibility of Weyl ordering. Thereby, for a quantum-mechanical system subject to an external potential $V(x)$, Schr\"{o}dinger's equation without ambiguity is
\begin{align}\label{Eq5}
    \bigg[\frac{1}{2m(x)}\hat{p}^2+\frac{i\hbar}{2}\frac{m'(x)}{m^2(x)}\hat{p}+V(x)\bigg]\Psi(x)=E\Psi(x).
\end{align}
Therefore, to avoid ambiguity arising from the symmetrization of the kinetic energy operator, allow us to consider Schrödinger's equation exposed in Eq. (\ref{Eq5}). So, by analyzing Eq. (\ref{Eq5}) with the results presented in the literature, we conclude that our theory is equivalent to the ordered BenDaniel-Duke PDM problem \cite{BenDaniel}. The BenDaniel-Duke theory has proven to be suitable for studying nanostructures subject to impurities, displacements, and geometric imperfections \cite{Willatzen}. It is essential to highlight that our manuscript will focus on describing the quantum eigenstates; however, there are other works examining issues on the PDM \cite{Rev1,Rev2,Rev3,Rev4,Rev5,Rev6,Rev7,Rev8}. For instance, one employs the PDM concept to investigate singular oscillator systems, viz., in which singularities arise due to singular interaction \cite{Rev1}. Furthermore, we use the PDM problems to describe several physical situations, e.g., transport of charge carriers in semiconductors with non-uniform composition \cite{Rev2}, and theoretical measures of quantum information \cite{Rev5}. Moreover, algebraic methods have shown great promise in the description of quantum eigenstates \cite{Rev3}.

\section{The singular PDM}\label{secIII}

The singular mass emerges in various contexts, allowing for its interpretation from different perspectives. One exciting scenario for singular mass occurs in high-energy physics. In this framework, singular mass refers to a point in space where the gravitational field becomes infinitely intense, such as at the center of a black hole, where one believes that matter collapses into a singularity, resulting in infinite density \cite{Carroll, Maluf}. Furthermore, the theory of singular mass also manifests in the fundamental principles of quantum mechanics \cite{Mustafa00}. In the quantum framework, singular mass is described through the motion of particles within the mass domain $m(x)$ \cite{Mustafa00}. Additionally, in quantum mechanics systems, a possible interpretation for PDM in quantum mechanical systems is regarded as a topological defect in the crystal lattice. Thus, one can understand the singular mass as a defect in the crystal lattice \cite{FLima}.

\begin{figure}[!ht]
    \centering
    \includegraphics[height=10cm,width=10cm]{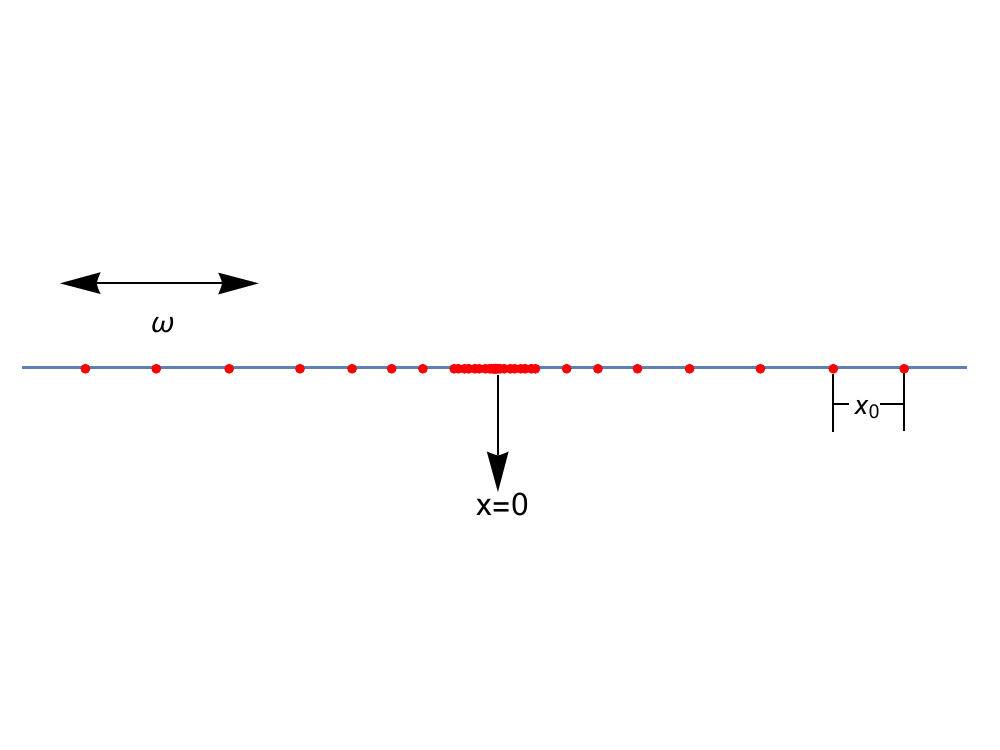}
    \vspace{-3.5cm}
    \caption{One-dimensional crystalline lattice with a singular mass distribution.}
    \label{LFig0}
\end{figure}

In our study, we are interested in exploring the singular PDM physical implications of this system within a non-relativistic one-dimensional theory. Thus, let us consider a one-dimensional crystalline lattice as illustrated in Fig. \ref{LFig0}. 

The PDM behaves as a lattice defect, inducing a continuous mass spectrum at $x=0$. One highlights that each atom (red dots) is subject to a restoring force, thus behaving quantum harmonic oscillators. Moreover, $x_0$ represents the equilibrium spacing between neighboring atoms.
\vspace{-4cm}

\begin{figure}[!ht]
    \centering
    \includegraphics[height=6.5cm,width=7.5cm]{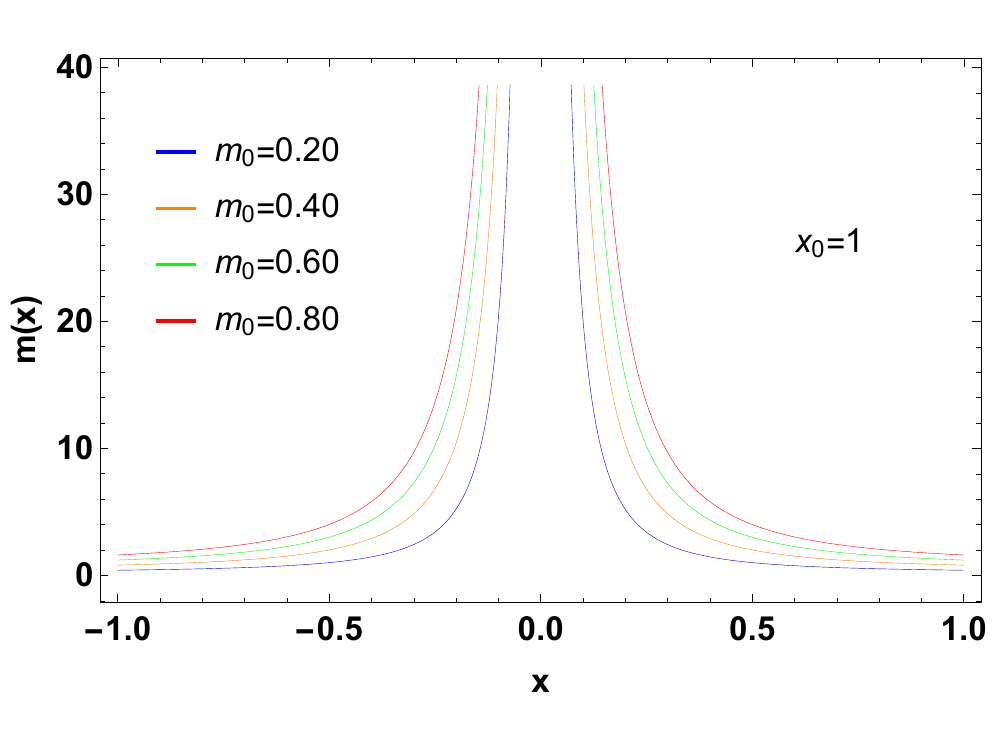}
    \includegraphics[height=6.5cm,width=7.5cm]{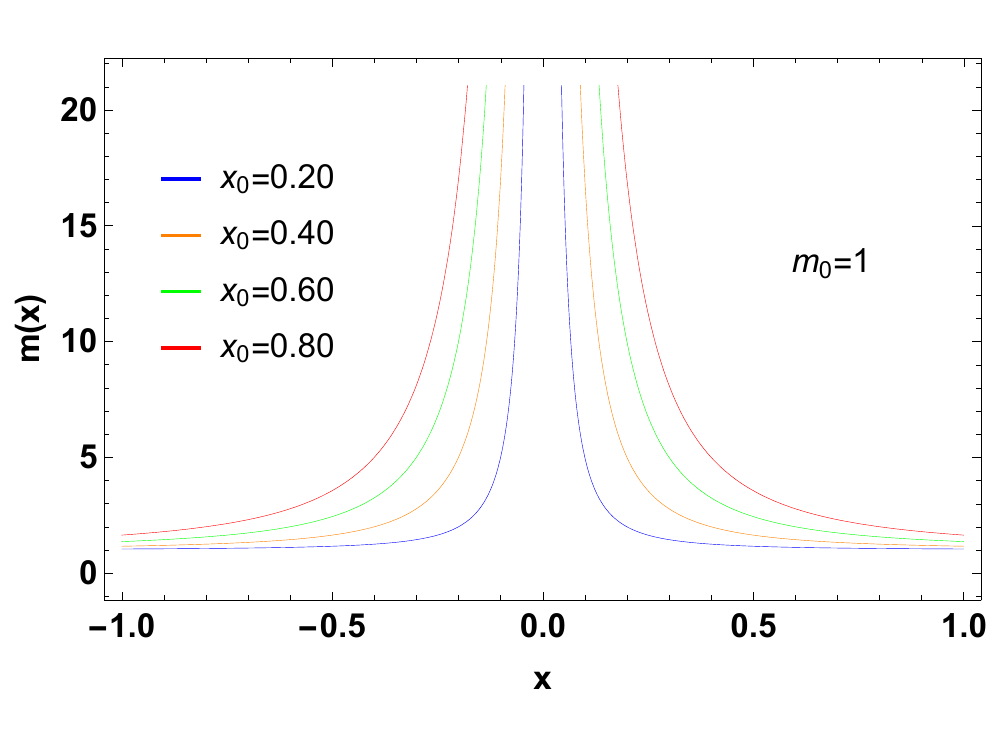}
    \vspace{-0.5cm}
    \begin{center}
     \hspace{0.5cm}   (a) \hspace{7cm} (b)
    \end{center}
    \vspace{-0.5cm}
    \caption{The behavior of the PDM vs. position. (a) PDM varying $m_0$ and keeping $x_0=1$. (b) PDM varying $x_0$ and keeping $m_0=1$.}
    \label{LFig1}
\end{figure}

Mathematically, we described our system using the PDM, namely,
\begin{align}\label{Eq6}
    m(x)=m_0\bigg(1+\frac{x_0^2}{x^2}\bigg).
\end{align}
Here, $m_0$ has mass dimensions and describes the intensity of the effective mass. Furthermore, $x_0$ represents the equilibrium spacing between lattice sites. We present the PDM varying $m_0$ (keeping $x_0$ constant) and varying $x_0$ (keeping $m_0$ constant), respectively, in Fig. \ref{LFig1}[(a) and (b)]. It is crucial to highlight that one can find some studies on the singular mass profiles in the literature \cite{Vakarchuk,ADAlhaidari,SMIkhdair}. For instance, the singular PDM can describe an electron confined by electromagnetic fields. Indeed, in this framework, this mass profile arises due to the multipole expanse, see Ref. \cite{Vakarchuk}.

\subsection{Connection with solid-state physics}

Allow us to attempt to establish a phenomenological connection with our model. To accomplish our purpose, let us recall that the PDM has an equivalent distribution in reciprocal space. Thereby, the PDM at the reciprocal space \footnote{One defines the mass at reciprocal space as Fourier's transform of the PDM at position space \cite{FLima}. The calculation of mass in momentum space, within the context of solid-state physics, is significant, as this approach facilitates the understanding of the relationships between the electronic and mechanical properties of materials and their crystal structure \cite{Kittel}. Furthermore, in momentum space, the dispersion energies of electrons are expressed as a function of the wave vector and the effective mass of charge carriers \cite{Kittel}.} is
\begin{align}\label{FEEq1}
    m(k)=\frac{1}{\sqrt{2\pi}}\int_{-\infty}^{+\infty}\,m(x)\,\text{e}^{-i k x}\,dx.
\end{align}

\begin{figure}[!ht]
    \centering
    \includegraphics[height=6cm,width=7.5cm]{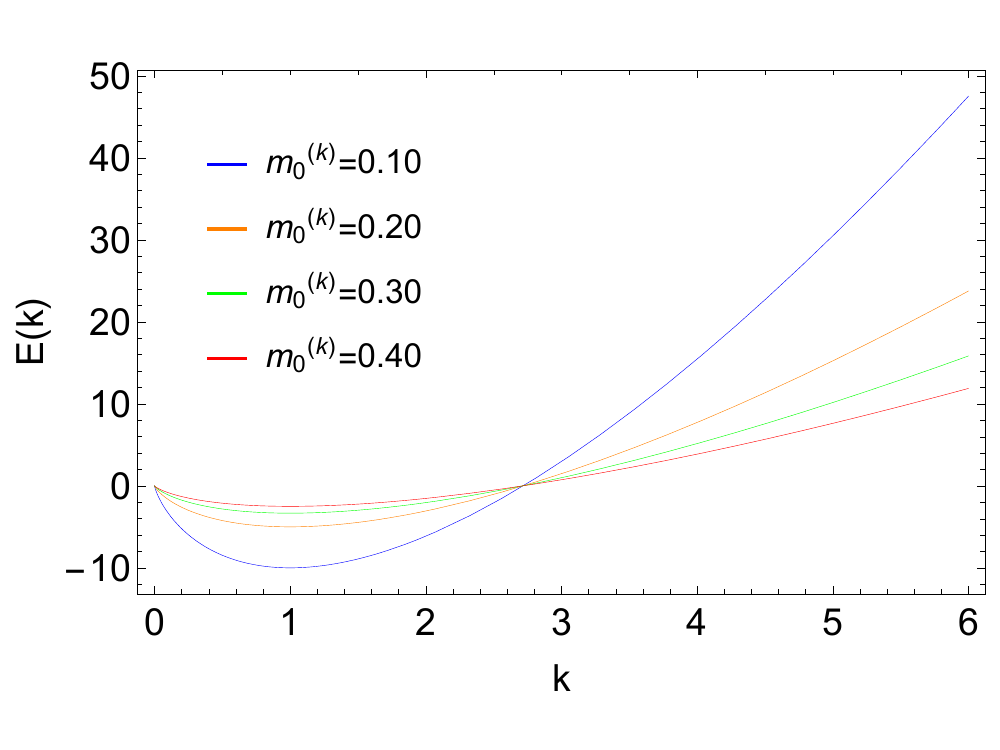}
    \vspace{-0.9cm}
    \caption{The dispersion energy concerning PDM [Eq. (\ref{Eq6})] with $\sigma=\hbar=1$.}
    \label{LFig2}
\end{figure}

Using the residue theorem and considering that the system's dispersion energy is
\begin{align}\label{FEEq2}
    \frac{d^2\textrm{E}(k)}{dk^2}=\frac{\hbar^2}{m(k)},
\end{align}
one obtains the dispersion energy, i.e., 
\begin{align}\label{FEEq3}
    \textrm{E}(k)=\frac{\hbar^2}{m_{0}^{(k)}}[k\, \text{ln}(\sigma k)-k].
\end{align}
Here, $m_{0}^{(k)}=\sqrt{2\pi}m_0 x_{0}^{2}$ and $\sigma$ is the jump parameter, which originates from the integration constant of the theory, ensuring the logarithmic argument dimensionless. Meantime, the parameter $m_0^{(k)}$ relates to the particle's energy (or electron) subjected to harmonic interaction as it propagates to neighboring points at the lattice. This propagation can extend closer to or farther from the singular mass region. This behavior of the crystalline structure suggests the formation of bound states in the PDM theory. Furthermore, the mass profile adopted is suitable for describing superconductivity phenomena if the parameter $m_0^{(k)}$ decreases. That is because, with a smaller $m_0^{(k)}$, the electrons in the lattice will be closer together, and the energy gap between them will be smaller. One highlights that our hypothesis is valid for theories described by the dispersion energy in Eq. (\ref{FEEq3}). We expose the dispersion energy in Fig. \ref{LFig2}.

\subsection{The eigenstates}

Let us turn our attention back to the quantum description of the eigenstates. To achieve our objective, we consider the Eq. (\ref{Eq5}) and the PDM (\ref{Eq6}). Thereby, one obtains Schrödinger's equation 
\begin{align}\label{Eq7}
    -\frac{\hbar^2}{2m_0}\Psi''(x)-\frac{\hbar^2}{m_0x^3}\bigg(1+\frac{x_0^2}{x^2}\bigg)^{-1}\Psi'(x)+\bigg(1+\frac{x_0^2}{x^2}\bigg)[V(x)-E]\Psi(x)=0.
\end{align}

Naturally, one assumes that the atoms in the crystalline lattice suffer the action of a restoring force \cite{Gomes,Kittel}, i.e., the atoms belonging to the lattice behave as a set of harmonic oscillators. Consequently, a PDM at a crystalline lattice will submitted to a restoring force, i.e., a harmonic-like interaction. Hence, the potential will be
\begin{align}\label{Eq8}
    V(x)=\frac{1}{2}m_0\, \omega_0^2 x^2.
\end{align}
One can find the behavior of the interaction $V(x)$ presented in Fig. \ref{LFig2}[(a) and (b)].
\begin{figure}[!ht]
    \centering
    \includegraphics[height=6.5cm,width=7.5cm]{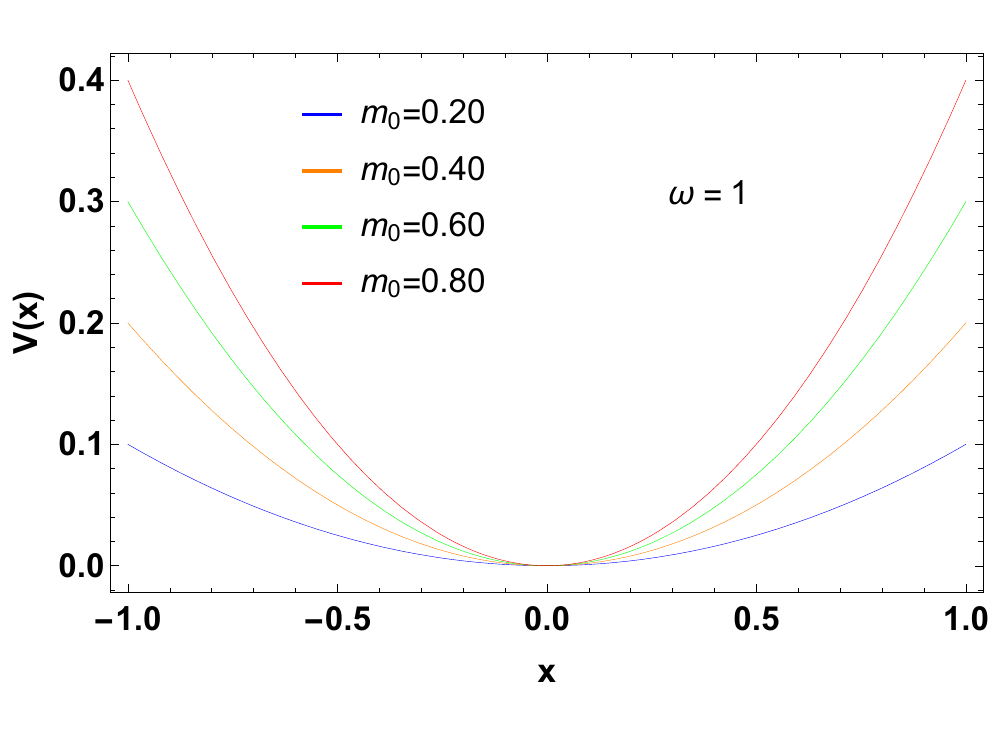}
    \includegraphics[height=6.5cm,width=7.5cm]{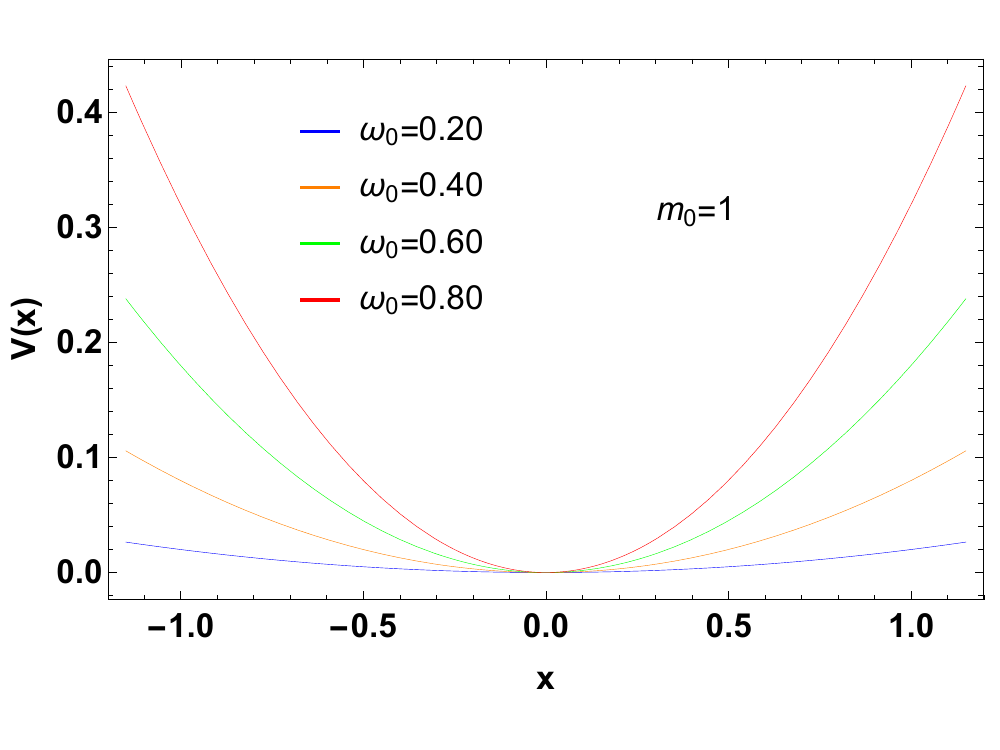}
    \vspace{-0.5cm}
    \begin{center}
     \hspace{0.5cm}   (a) \hspace{7cm} (b)
    \end{center}
    \vspace{-0.5cm}
    \caption{Harmonic interaction of the quantum mechanical system. (a) Potential varying $m_0$ keeping $\omega_0$ constant. (b) Potential varying $\omega_0$ and keeping $m_0$ constant.}
    \label{LFig3}
\end{figure}

Considering a system subjected to a harmonic interaction, one applies the change of coordinates 
\begin{align}\label{Eq9}
    x\to\sqrt{\frac{\hbar}{m_0\,\omega_0}}\xi,
\end{align}
This variable's change is reported in the literature \cite{Sakurai}. Here, $x$ and $\xi$ varies in the range $[-\infty,+\infty]$. Accordingly, adopting the change of variable (\ref{Eq9}), we arrive to
\begin{align}\label{Eq10}
    \Psi''(\xi)+\bigg[\frac{2  \Gamma}{\xi (\xi^2+\Gamma)}\bigg]\, \Psi'(\xi)+(\mathcal{E}-\xi^2)\bigg(1+\frac{\Gamma}{\xi^2}\bigg)\Psi(\xi)=0. 
\end{align}
In this case, one defines the parameters $\Gamma=m_0 \omega x_{0}^{2}/\hbar$ and $\mathcal{E}=2E/\hbar\omega$.

Now, allow us to assume the dependent coordinate change $\Psi(\xi)=\text{e}^{-\frac{\xi^2}{2}}h(\xi)$, this leads us to
\begin{align}\label{Eq11}
    h''(\xi)+2\xi\bigg[\frac{\Gamma}{\xi^2(\xi^2+\Gamma)}-1\bigg]h'(\xi)+\bigg[\mathcal{E}-\Gamma-1-\frac{2\Gamma}{\xi^2+\Gamma}+\frac{\mathcal{E}\Gamma}{\xi^2}\bigg]h(\xi)=0.
\end{align}
In the case $\Gamma=0$, we recover the analytical equation of the usual harmonic oscillator, i.e., constant mass distribution, namely,
\begin{align}\label{Eq12}
    h''(\xi)-2\xi h'(\xi)+(\mathcal{E}-1)h(\xi)=0.
\end{align}
In this regime, the system's solutions are Hermite's polynomials. In summary, this occurs because if $\Gamma\to 0$, the PDM converges to the problem of a particle with constant mass confined to a harmonic interaction. Thus, $m(x)\to m_0$ under the action of the interaction (\ref{Eq8}).

Appropriately, we assume the change $\xi^2\to z$. That enables us to obtain
\begin{align}\label{Eq13}
    4zh''(z)+4z\bigg[\frac{  \Gamma}{z(z+\Gamma)}+\frac{1/2}{z}-1\bigg]h'(z)+\bigg(\mathcal{E}-1-\Gamma-\frac{2\Gamma}{z+\Gamma}+\frac{\mathcal{E}\Gamma}{z}\bigg)h(z)=0. 
\end{align}
where $z\in (0,\infty]$. By studying the equation (\ref{Eq13}), one considers the transformation $z\to-\Gamma y$ to shift the singularity around the rest mass. This approach allows us to write
\begin{align}\label{Eq14}
    h''(y)+\bigg[\Gamma+\frac{3/2}{y}-\frac{1}{y-1}\bigg]h'(y)-\frac{\Gamma}{4y}\bigg[\mathcal{E}-\Gamma-1+\frac{2}{y-1}-\frac{\mathcal{E}}{y}\bigg]h(y)=0.
\end{align}

To achieve our purpose and demonstrate the analytical solutions of the system, let us apply one last transformation, i.e., $h(y)=y^\lambda H(y)$ with $\lambda=-\frac{1}{4}\pm\frac{1}{4}\sqrt{1-4\mathcal{E}\Gamma}$. This transformation leads us to the
\begin{align}\label{Eq15}
    H''(y)+\bigg[\Gamma+\frac{3/2+2\lambda}{y}-\frac{1}{y-1}\bigg]H'(y)+\bigg[\frac{\lambda(\Gamma+1)+\frac{\Gamma}{2}(1-\nu/2)}{y}-\frac{\lambda+\frac{\Gamma}{2}}{y-1}\bigg]H(y)=0.
\end{align}
The equation (\ref{Eq15}) is called the confluent Heun equation. For further details, see Ref. \cite{Ronveaux}. By comparing Eq. (\ref{Eq15}) with the confluent Heun equation, one concludes that the solutions of Eq. (\ref{Eq15}) are
\begin{align}\label{Eq16}
    H^{(1)}(y)=\text{Hc}(\alpha,\,\beta,\,\gamma,\,\delta,\, \eta;\, y),
\end{align}
and
\begin{align}\label{Eq17}
    &H^{(2)}(y)=y^{-\beta}\text{Hc}(\alpha,\,-\beta,\, \gamma,\,\delta,\,\eta;\, y).
\end{align}
The functions $H^{(1,2)}(y)$ are linearly independent. Furthermore, one highlights that for these wave functions to be physically acceptable, one must satisfy the conditions
\begin{align}\label{Eq18}
    \frac{\delta}{\alpha}+\frac{\beta+\gamma}{2}+1=-n,
\end{align}
and 
\begin{align}
    \label{EEq18}
    \Delta_{n+1} (\mu)=0.
\end{align}
Here, $\mu=\frac{1}{2}(\alpha-\beta-\gamma+\alpha\beta-\beta\gamma)-\eta$ \footnote{It is pertinent to highlight that these conditions ensure that the eigenfunctions and their respective derivatives are continuous at the origin (i.e., $x=0$).}. For further details, see Ref. \cite{Fiziev}.

By algebraically inspecting Eqs. [(\ref{Eq18}) and (\ref{EEq18})] and considering the confluent Heun equation presented in Refs. \cite{Ronveaux, Fiziev}, one concludes that
\begin{align}\label{Eq19}
    &\alpha=\Gamma,\\ \label{Eq19.2}
    &\beta=\pm\frac{1}{2}\sqrt{1-4\mathcal{E}\Gamma},\\ \label{Eq19.3}
    &\gamma=-2,\\ \label{Eq19.4}
    &\delta=-\frac{\Gamma}{4}(\mathcal{E}-\Gamma),\\ \label{Eq19.5}
    &\eta=\frac{4(\Gamma+1)-\mathcal{E}\pm3\sqrt{1-4\mathcal{E}\Gamma}}{4}.
\end{align}

Substituting Eqs. [(\ref{Eq19})-(\ref{Eq19.5})] into Eqs. [(\ref{Eq18}) and (\ref{EEq18})], one obtains the energy spectrum, namely,
\begin{align}\label{Eq20}
    E_n=\bigg(2n-\frac{\Gamma}{2}\pm\frac{1}{2}\sqrt{1-16n\Gamma}\bigg)\hbar\omega_0 \hspace{0.65cm} \text{with} \hspace{0.65cm} n=0,1,2,\dots 
\end{align}
Note that in the limit $m(x)\to m_0$, we reach the spectrum of the standard harmonic oscillator, i.e.,
\begin{align}\label{0Eq20}
    E_n=\bigg(n+\frac{1}{2}\bigg )\hbar\omega_0 \hspace{0.65cm} \text{with} \hspace{0.65cm} n=0,1,2,\dots 
\end{align}
The energy spectrum imposes the constraint
\begin{align}
    \Gamma\leq \frac{1}{16n},
\end{align}
i.e.,
\begin{align}\label{FreqEq}
    \omega_0\geq \frac{\hbar}{16n m_0x_{0}^{2}} \hspace{0.65cm} \hspace{0.75cm} \text{with} \hspace{0.75cm} n=1,2,3,\dots
\end{align}
Therefore, the harmonic oscillator with singular mass has quantized frequency. Here, one highlights that the quantization (or constraint) that arises in the frequency \eqref{FreqEq} is a consequence of the existence singularity in the mass distribution at $x = 0$. Thus, this restriction is directly related to the energy's real eigenvalues.

\begin{figure}[!ht]
    \centering
    \includegraphics[height=6.5cm,width=7.5cm]{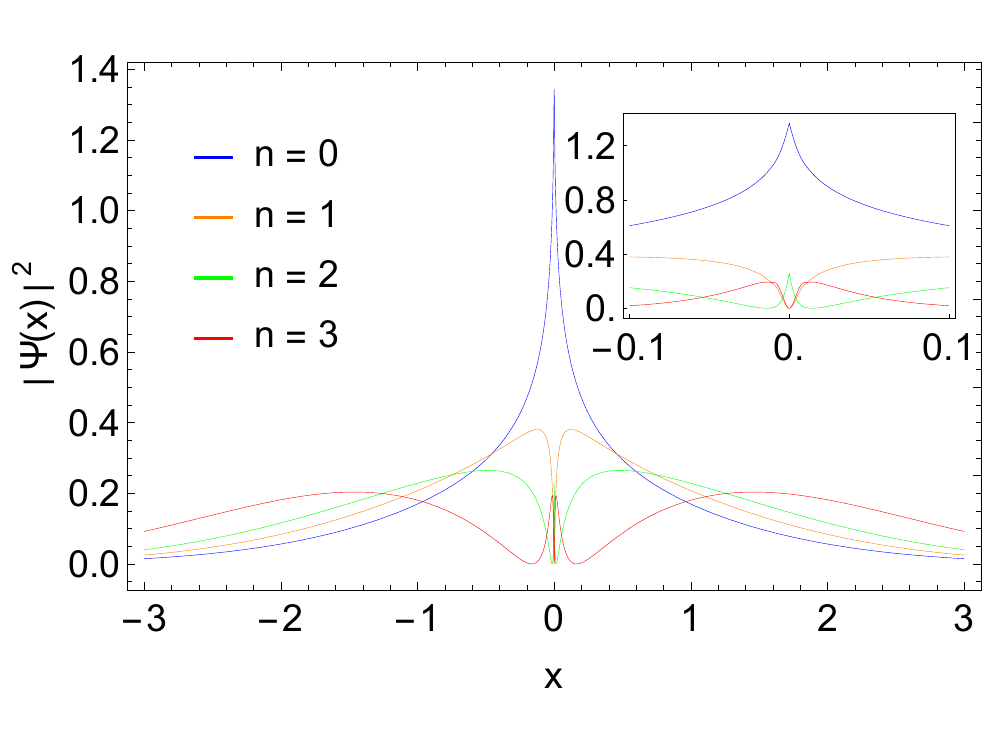}
    \includegraphics[height=6.5cm,width=7.5cm]{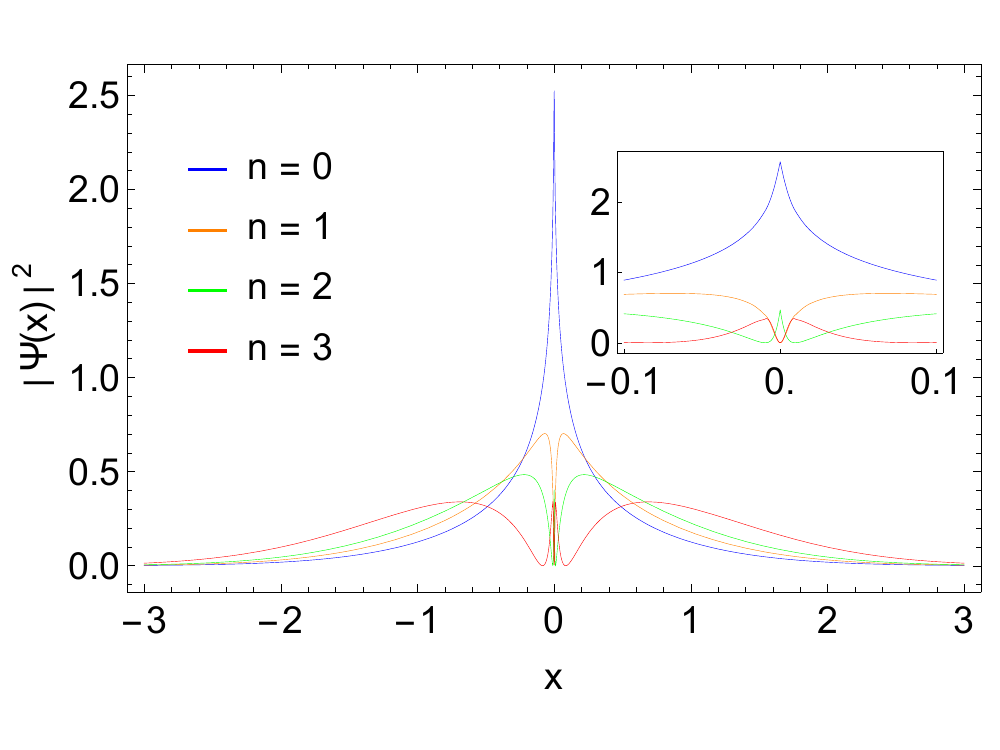}
    \vspace{-0.5cm}
    \begin{center}
     \hspace{0.5cm}   (a) \hspace{7cm} (b)
    \end{center}    \vspace{-0.5cm}
    \includegraphics[height=6.5cm,width=7.5cm]{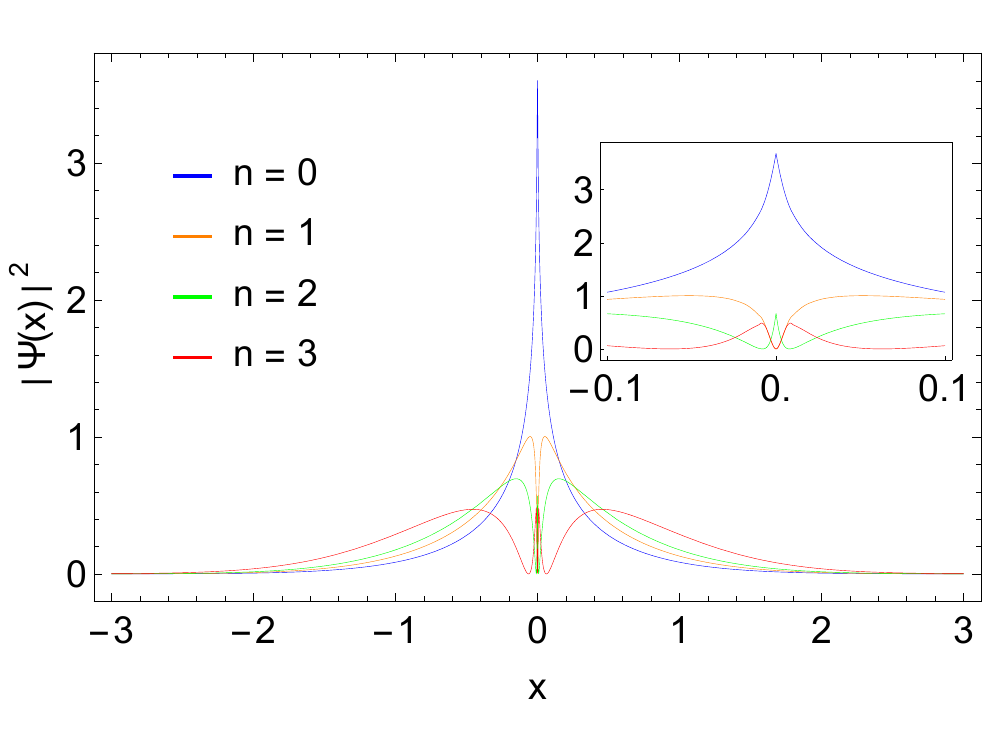}
    \includegraphics[height=6.5cm,width=7.5cm]{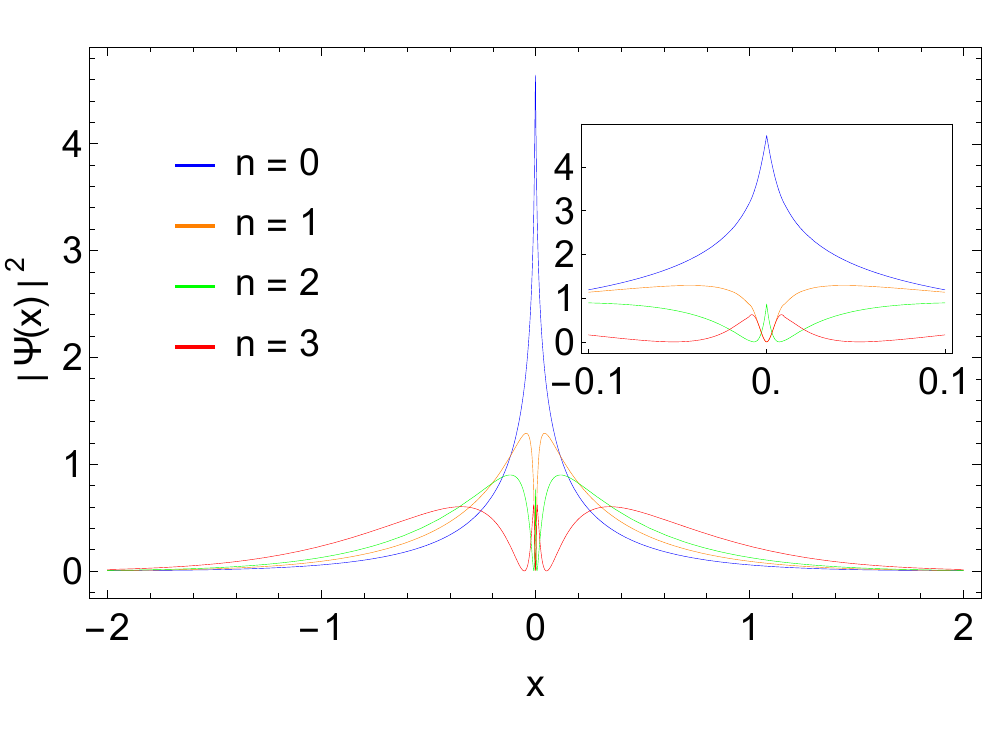}
    \vspace{-0.5cm}
    \begin{center}
     \hspace{0.5cm}   (c) \hspace{7cm} (d)
    \end{center} 
    \vspace{-0.5cm}
    \caption{Probability density concerning the solutions of the Eqs. [(\ref{Eq21}) and (\ref{Eq22})]. Figures (a), (b), (c), and (d) represent the probability densities when the frequency is $0.2$, $0.4$, $0.6$, and $0.8$.}
    \label{LFig44}
\end{figure}

In terms of the position variable $x$, the wave eigenfunctions of Eq. (\ref{Eq7}) are
\begin{align}\label{Eq21}
    \Psi^{(1)}(x)=\bigg(-\frac{x^2}{x_{0}^{2}}\bigg)^\lambda\text{e}^{-\frac{m_0\omega x^2}{2\hbar}}\,\text{Hc}\bigg(\alpha,\, \beta,\, \gamma,\,  \delta,\, \eta;\,-\frac{x^2}{x_0^2}\bigg),
\end{align}
and
\begin{align}\label{Eq22}
        \Psi^{(2)}(x)=\bigg(-\frac{x^2}{x_{0}^{2}}\bigg)^{(\lambda-\beta)}\text{e}^{-\frac{m_0\omega x^2}{2\hbar}}\,\text{Hc}\bigg(\alpha,\, -\beta,\, \gamma,\,  \delta,\, \eta;\,-\frac{x^2}{x_0^2}\bigg).
\end{align}
Heun's functions $\Psi^{(1,2)}(x)$ constitute the complete set of solutions for the quantum system. We expose in Fig. \ref{LFig44} the plots of these eigenfunctions (the corresponding probability densities) for the first four-energy eigenstates when the frequency varies.

\section{Final remarks}\label{secIV}

We addressed the issue of a generic PDM. Strictly speaking, one revisits the problem of the non-relativistic kinematic operator. Besides, we considered the von Roos ordering \cite{Roos} supplemented by the kinematic term $\hat{K}_0$, viz., 
\begin{align}
    \hat{K}_0=\frac{1}{4}\bigg[\frac{1}{m(r)}\hat{p}^2+\hat{p}^2\frac{1}{m(r)}\bigg].    
\end{align}
The term $\hat{K}_0$ leads us to theories of non-uniform semiconductors with position-dependent chemical composition. Furthermore, one highlights that our theory preserves the properties of von Roos \cite{Roos} ordering and selects a single appropriate ordering for the Hamiltonian with PDM, i.e., BenDaniel-Duke Hamiltonian \cite{BenDaniel}.

Adopting Schrödinger's theory with BenDaniel-Duke Hamiltonian \cite{BenDaniel}, one describes a one-dimensional crystalline lattice with a singular PDM defect. In this conjecture, we noted that the one-dimensional crystalline lattice tends to exhibit an almost atom's continuous distribution (or a cluster) near the singularity of the effective mass. Hence, if each crystalline lattice site is characterized by oscillating atoms, one can treat the effective system according to the Schrödinger theory [Eq. (\ref{Eq5})], subject to a harmonic impurity (or interaction) [Eq. (\ref{Eq8})]. Considering this, we investigated the system's quantum eigenstates. Thus, one concludes that the model's wavefunctions are Heun's functions with eigenvalues reported by Eq. (\ref{Eq20}). Nevertheless, when $m(x) \to m_0$, i.e., the PDM problem reduces to a particle with constant mass subjected to a harmonic potential, we noted Hermite's polynomials, which lead us to describe the quantum eigenstates with energy
\begin{align}
    E_n=\bigg(n+\frac{1}{2}\bigg)\hbar\omega_0 \hspace{0.5cm} \text{with} \hspace{0.5cm} n=0,1,2,\dots
\end{align}
An analysis of the probability densities $\vert\Psi^{(1,2)}(x)\vert^2$ described by confluent Heun functions reveals that the regions of highest probability of locating the particle cluster reside at the vicinity of $x=0$. However, one notes that this probability density tends to decrease at higher energy levels. This phenomenon arises from the widening of uncertainties in the position of the singular PDM at higher energy levels. That result is due to the information loss about the particle's quantum state. For a more detailed analysis of quantum information measures of a PDM, see Ref. \cite{FLima}.

\section{Acknowledgment}

The authors wish to express their gratitude to FAPEMA and CNPq (Brazilian research agencies) for their invaluable financial support. F. C. E. L. is supported by FAPEMA BPD-05892/23. F. M. B. is supported by CNPq 161092/2021-7. C. A. S. A. is supported by CNPq 309553/2021-0, CNPq/Produtividade. Furthermore, C. A. S. A. is grateful to the CNPq, project number 200387/2023-5, and acknowledges the Department of Theoretical Physics \& IFIC from the University of Valencia for their warm hospitality. 

\section{Conflicts of Interest/Competing Interest}

The authors declared that there is no conflict of interest in this manuscript.

\end{document}